\documentclass[twocolumn]{aastex62}

\received{}
\revised{}
\accepted{}
\submitjournal{ApJ}

\shorttitle{X-ray emission from LMC-N57}
\shortauthors{Ramirez-Ballinas et al.}

\begin{document}

\title{Dissecting the hot bubbles in LMC-N57 with {\it XMM-Newton}}

\correspondingauthor{Isidro\,Ramirez-Ballinas}
\email{iramirez@astro.unam.mx}
\author{Isidro Ram\'irez-Ballinas}
\affil{Instituto de Astronom\'{\i}a, Universidad Nacional
  Auton\'oma de M\'exico, Apdo. Postal 70-264, 04510 Mexico
City, Mexico}

\author{Jorge\,Reyes-Iturbide}
\affil{Divisi\'on de Mec\'anica, Tecnol\'ogico de Estudios Superiores de
  Tianguistenco, Carretera Tenango La Marquesa Km 22, Santiago
  Tianguistenco, Estado de M\'exico, Mexico}

\affil{Instituto de Ciencias Nucleares, 
Universidad Nacional Aut\'onoma de M\'exico, Apdo. Postal 70-543, 04510, Mexico City, Mexico}

\author[0000-0002-5406-0813]{Jes\'us A.\,Toal\'a }
\affiliation{Instituto de Radioastronom\'{\i}a y Astrof\'{i}sica, UNAM Campus Morelia, Apartado postal 3-72, 58090 Morelia, Michoac\'{a}n, Mexico}

\author[0000-0003-1113-2140]{Margarita\,Rosado}
\affil{Instituto de Astronom\'{\i}a, Universidad Nacional
  Auton\'oma de M\'exico, Apdo. Postal 70-264, 04510 Mexico
City, Mexico}




\begin{abstract}

We present a study of the diffuse X-ray emission from the star forming
region LMC-N\,57 in the Large Magellanic Cloud (LMC). We use archival
{\it XMM-Newton} observations to unveil in detail the distribution of
hot bubbles in this complex. X-ray emission is detected from the
central superbubble (SB) DEM L\,229, the supernova remnant (SNR)
0532$-$675 and the Wolf-Rayet (WR) bubble DEM L\,231 around the WR
star Br\,48. Comparison with infrared images unveils the powerful
effect of massive stars in destroying their nurseries. The
distribution of the hot gas in the SNR and the SB display their maxima
in regions in contact with the filamentary cold material detected by
IR images. Our observations do not reveal extended X-ray emission
filling DEM L\,231, although several point-like sources are detected
in the field of view of this WR nebula. The X-ray properties of Br\,48
are consistent with a binary WN4$+$O as proposed by other authors. We
modelled the X-ray emission from the SB and found that its X-ray
emission can be simply explained by pressure-driven wind model, that
is, there is no need to invoke the presence of a SN explosion as
previously suggested. The pressure calculations of the hot gas
confirms that the dynamical evolution of the SB DEM L\,229 is
dominated by the stellar winds from the star cluster LH\,76.

\end{abstract}




\section{Introduction} \label{sec:intro}

\begin{figure*}
\begin{center}
\includegraphics[width=0.5\linewidth]{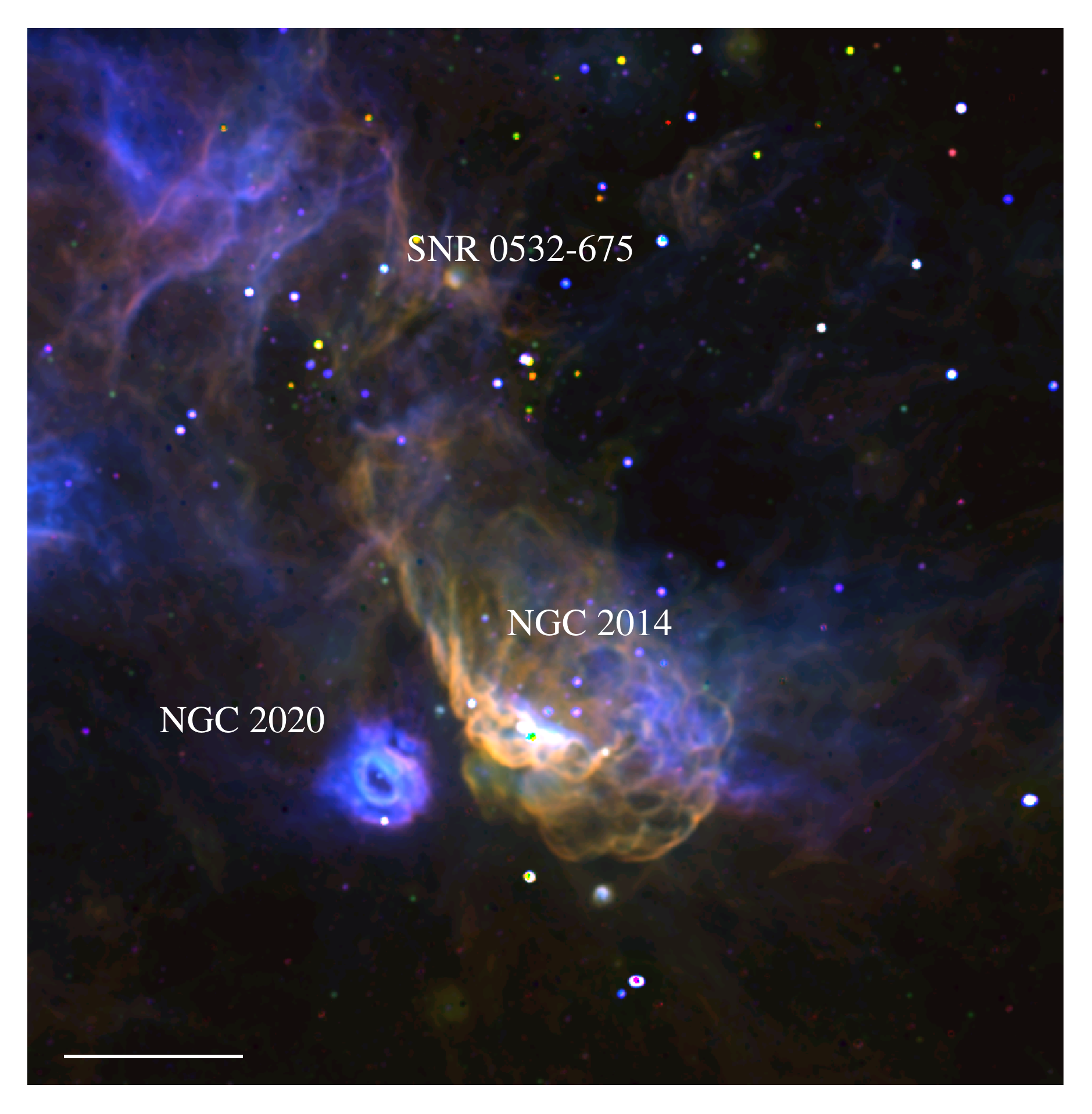}~
\includegraphics[width=0.5\linewidth]{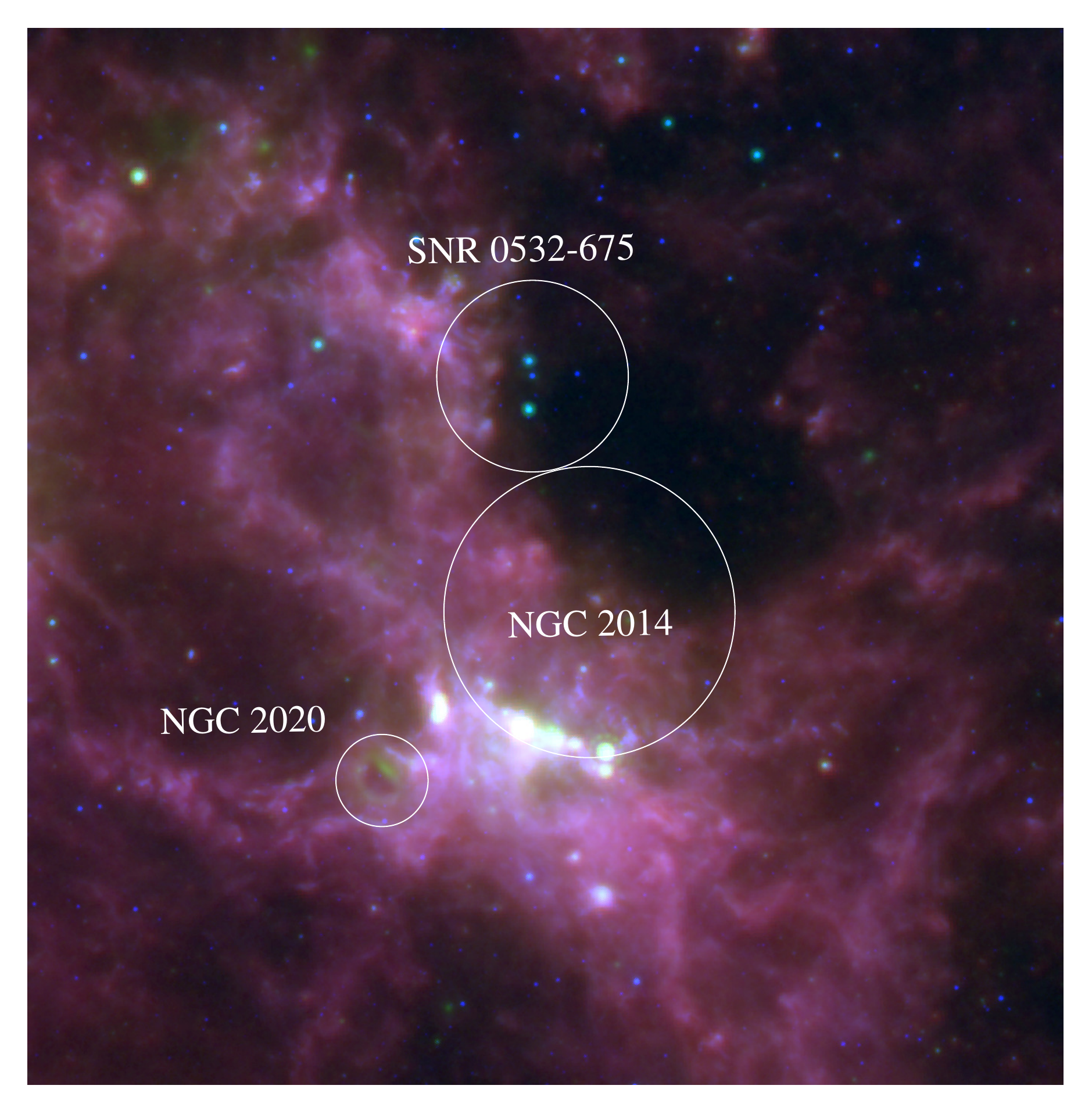}
\caption{Optical and IR colour-composite images of LMC-N\,57. Left:
  Optical image obtained with the MCELS survey \citep{Smith2005}. Red,
  green and blue correspond to [S\,{\sc ii}], H$\alpha$, and [O\,{\sc
      iii}], respectively. Right: IR image. Red, green, and blue
  correspond to {\it Herschel} SPIRE 250~$\mu$m, {\it Spitzer} MIPS
  24~$\mu$m, and {\it Spitzer} IRAC 8.0~$\mu$m. The white bar
  corresponds to 5~arcmin (72~pc) and both images have the same field
  of view. North is up and east to the left.}
\label{fig:rgb1}
\end{center}
\end{figure*}

During their lives, massive stars
($M_\mathrm{i}\gtrsim$10~M$_{\odot}$) drive the evolution of the
interstellar medium (ISM) via energy and momentum. This feedback is
mainly due to their stellar winds, radiation pressure and supernova
(SN) explosions \citep{krum14}. At large scales ($\geq$1~kpc), stellar
UV radiation from young hot stars play a crucial role in the evolution
galaxies by regulating star formation \citep{ost}. Stellar feedback is
a necessary effect for formation and evolution of galaxies in
numerical simulations. This has been found to prevent the "cooling
catastrophe" \citep{ker}, the cause of low stellar density and stellar
population gradients observed in dwarf galaxies \citep{mas}.

At smaller scales ($<$100~pc), stellar feedback of star clusters
influence their surrounding giant molecular clouds. This stellar
feedback is an important parameter in explaining why only a small
fraction of the giant molecular cloud mass is converted into stars,
whilst driving the dynamic of the region. Massive stars dissipate the
surrounding molecular cloud by the combination of their powerful
stellar winds and the extra contribution from SN explosions, creating
the so-called bubbles and superbubbles \citep[SB;][]{chum90, chu95,
  coop04, oey96a, oey96b, ros86, george83, dunne01,
  reyes14,zhang14,re09,ve13,ary11}.

Numerical simulations have shown that both effects, ionization photon
flux and winds, have to be taken into account to produce a more
accurate study of the dynamics of bubbles around massive stars
\citep[e.g.,][]{toa11, dw13, fr06}. In those cases, the hot bubble
inside the H\,{\sc ii} region still dominates the dynamics of the
nebulae around massive stars. However, there are some observational
works that suggest otherwise. For example, \cite{lopez} analyzed
multi-wavelength data to test the driving mechanisms of SBs in the
Magellanic Clouds (MCs). These authors estimate different pressure
contributions from stellar radiation ($P_\mathrm{dir}$), the
dust-processed radiation ($P_\mathrm{IR}$), the warm ionized gas
($P_\mathrm{HII}$) and that coming from the adiabatically-shocked hot
bubbles ($P_\mathrm{X}$). These authors found that the warm ionized
medium dominates over the rest of the feedback processes. So that the
warm ionized gas plays a major role in the dynamics of star forming
regions. \citet{lopez} found a characteristic radius $r_\mathrm{ch}$
where the given source transitions from radiation pressure drive the
gas pressure driven by setting the total radiation pressure equal to
the warm gas pressure. This suggests that the radiation pressure is
important in the range $r_\mathrm{ch}\sim [0.01-7]$~pc. The presented
sample of analyzed SBs have radii between $\sim 10-50~\rm pc$ for
which they are too large to the radiation pressure dominates in the
dynamic of the shells from HII regions. We note that \citet{lopez}
studied the properties of the hot gas in the star forming regions in
the MCs by analyzing {\it ROSAT} X-ray observations. Although this
satellite had a good spectral sensitivity in the soft X-ray range
(where hot bubbles emit their bulk of X-ray emission) current X-ray
missions have a superior effective area and angular and spectral
resolution, which allows a better rejection of contaminant point
sources.

We have started a project to study the X-ray properties of star
forming regions in the MCs using {\it XMM-Newton} observations to peer
into this discrepancy. In this work we study the dynamical role of
stellar feedback mechanisms in the LMC-N\,57 complex (see
Fig.~\ref{fig:rgb1}). In particular, SBs in the Large Magellanic Cloud
(LMC) represent excellent cases of study due to its nearly face-on
disk (inclined by $10^{\circ}-30^{\circ}$ to the line of sight), and
because the interstellar reddening toward this galaxy is
small. Furthermore, the LMC provides a sample of more than 200 SBs
\citep[see, e.g., the original work of][]{Henize1956} at a common
distance $\sim 50~\rm kpc$ \citep{fea99} that are resolvable by
observations of current X-ray satellites.

LMC-N57 is composed by three main regions: a central superbubble (DEM
L\,229 or NGC\,2014), a Wolf-Rayet (WR) nebula (DEM L\,231 or
NGC\,2020) and the SNR\,0532$-$675 located north from the SB (see
Fig.~\ref{fig:rgb1}). The SB seems to have formed as the result of the
feedback from the OB association (OBA) LH\,76 \citep{lh70}. \cite{deg}
used optical photometry to determine that LH\,76 appears to be coeval
with an estimated age of $2-5\times10^{6}$~yr ago. Although this is a
sufficient time for the first supernova explosion to occur, radio
observations reported a spectral index $\alpha=0.1$ \citep{mc}, more
in agreement with H\,{\sc ii} region radio emission than the typical
values for a SNR. On the other hand, the WR nebula DEM L\,231 has a
"double-rim" morphology with a physical size of 24~pc$\times$17~pc
that seems to have been produced by the stellar wind of only one star,
the WR star Br\,48 \citep{Chu1999}.

Early analysis of X-ray observations of LMC-N\,57 obtained with the
{\it Einstein} observatory suggested at a relatively low X-ray
luminosity compared to other star forming complexes
\citep[][]{Wang1991}. For example, \citet{chum90} reported a
luminosity of $7.4\times0^{34}$~erg~s$^{-1}$ for LMC-N\,57 while other
complexes reached X-ray luminosities at least two orders of magnitude
higher. The analysis of {\it ROSAT} PSPC observations presented by
\citet{dunne01} showed that the SB is filled with hot gas, that the
SNR\,0532$-$675 is an X-ray bright source and that no extended
emission is detected from the WR bubble Br\,48 \citep[see also
  figure~7 in][]{Points2001}. The best-fit model to the PSPC spectrum
resulted in a plasma temperature of $kT=0.26$~keV
($T=2.6\times10^{6}$~K) and an X-ray luminosity in the 0.5--2.4~keV
energy range of $L_\mathrm{X}=[2.1-8.1]\times10^{35}$~erg~s$^
{-1}$. \citet{dunne01} compared their results with those of
pressure-driven models and concluded that the X-ray emission detected
from the 13 sources analyzed in that work was the result of other
physical processes such as SN explosions in addition to stellar
winds. We note, however, that their X-ray luminosity estimates for
LMC-N\,57 were very close to the observable ones.

In this paper we present the analysis of archival {\it XMM-Newton}
observations in comparison with optical and IR observations. The {\it
  XMM-Newton} view of LMC-N\,57 largely improves the previous X-ray
studies of this complex. This study unveils the detailed distribution
of the X-ray-emitting gas in LMC-N\,57. We compare our results with
the theoretical predictions from wind-blown bubble formation and X-ray
emission. This paper is organized as follows. In Section~2 we present
a description of the analytical model of SBs. Section~3 describes our
{\it XMM-Newton} observations. The distribution of the X-ray-emitting
gas is described in Section~4 and the analysis of its spectral
properties is presented in Section~5. Finally, the discussion and
conclusions are presented in Sections~6 and 7, respectively.

\section{Theoretical predictions of the X-ray emission from superbubbles}

The standard model of bubble formation by stellar winds was proposed
by \cite{w77}. Although the model considers a single wind source, it
has been extended to describe the structure and evolution of the SB
formed by several stars in a cluster \citep{chu95}.

The stars in the OBA deposit mechanical energy into the ISM through their stellar winds. This is given by
\begin{equation}
L_w = \sum_{i=1}^N \frac{1}{2} \dot{M}_{w,i} v^2_{w,i},
\label{lw}
\end{equation}

\begin{figure}[]
\begin{center}
\includegraphics[width=0.8\linewidth]{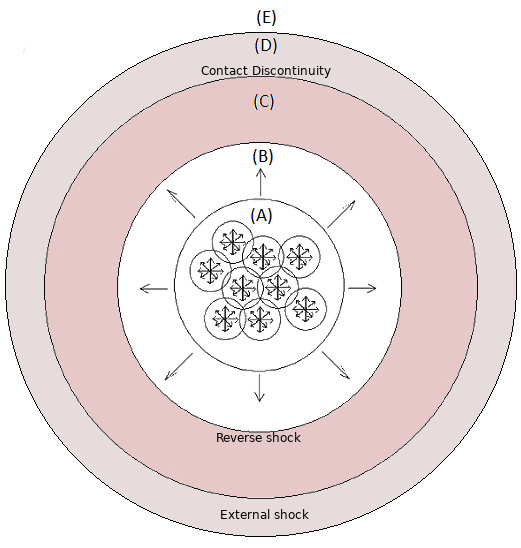}
\caption{Generalization of the standard model of wind-blown bubbles
  conceived for a single star. Schematic structure of a SB produced by
  an OBA: (A) star cluster zone, (B) free-wind zone, (C)shocked-wind
  zone, (D) shocked-ISM zone, and (E) unperturbed ISM.}
\label{fig0}
\end{center}
\end{figure}

\noindent where $\dot{M}_{w,i}$ and $v_{w,i}$ are the mass-loss rate
and the wind velocity of $i$-th star, respectively, and $N$ is the
total number of stars. The interaction of these winds with ISM creates
a SB structure with the following four zones (see Figure \ref{fig0}):

(A) a central zone where the stars are located and inject their
free-flowing winds. This region is delimited by the stellar cluster
radius. Outside this region a common cluster wind is established;

(B) a free wind zone is the region between the OBA radius and reverse
shock. This region is filled by the unperturbed stellar OBA wind;

(C) a zone of shocked cluster wind, located between the reverse shock
(or inner shock) and the contact discontinuity. This material has been
shock-heated and reaches temperature in excess to $10^{7}$~K that
emits primarily in X-rays;

(D) an external zone, between the contact discontinuity and the
leading shock. This zone contains the shocked ISM material that has
been swept by the leading shock with an important optical line
emission;

(E) the unperturbed ISM.

In zone (A), stellar winds of the massive stars collide with each
other, thermalizing all the gas injected inside the cluster volume
forming a common OBA wind, which produces an over-pressure inside the
cluster volume with respect to its environment.  The OBA wind expands
freely inside zone (B). Zone (C) is formed by gas of the stellar
cluster wind that has been shocked by the inner or reverse shock. This
zone is adiabatically-shocked and its post-shock temperature can be
estimated to exceed 10$^{7}$--10$^{8}$~K \citep{Dyson1997}. Finally,
zone (D) is formed by the shocked ISM gas that emits mainly in
optical, is the densest zone according to the standard model.

\cite{w77} showed that the equations that describe the dynamics of the
shell or zone (D) can be expressed as:

\begin{equation}
R_\mathrm{pc}= 42\,\mathrm{pc} \times L^{1/5}_{w37} n^{-1/5}_0 t^{3/5}_6,
\label{radius}
\end{equation}

\begin{equation}
V=\frac{dR}{dt}= 0.59\,\mathrm{km\,s}^{-1} \times R_\mathrm{pc}/t_6,
\label{veloc}
\end{equation}

\noindent where $R_\mathrm{pc}$ is the radius of the SB in pc, $V$ is
the expansion velocity of the SB, $L_{w37}$ is the mechanical
luminosity of the OBA in units of $10^{37}$ erg s$^{-1}$, and $n_0$
and $t_6$ are the number density of the ambient medium in units of
cm$^{-3}$ and the age of the bubble in $10^6$~yr, respectively.

For the case that the radius of the shell and its expansion velocity
are known, it is possible to determine the ambient density in terms of
wind luminosity $L_w$ (Eq.~\ref{lw}), radius, and bubble age by the
relation
\begin{equation}
n_0 = (1.3 \times 10^8 \mathrm{cm}^{-3}) L_{w37} t^3_\mathrm{Myr} R^{-5}_\mathrm{pc}.
\label{dens}
\end{equation}
 
According to \citet{w77}, thermal conductivity regulates the
temperature of the hot bubble toward its outer edge. The region in
contact with the swept ISM reduces its temperature to values
$\sim10^{6}$~K while raising its density, producing soft X-ray
emission. The X-ray luminosity that arises from the internal
conduction front can be estimated following \citet{chum90} as
\begin{equation}
L_\mathrm{X}=(1.1\times 10^{35}~\mathrm{erg~s^{-1}})I(\tau)\xi L_{w37}^{33/35}n_0^{17/35}t_6^{19/35},
\label{Lx}
\end{equation}
\noindent where $\xi$ is the gas metallicity, and $I(\tau)$ can be expressed as
\begin{equation}
I(\tau)=\frac{125}{33}-5\tau^{1/2}+\frac{5}{3}\tau^3-\frac{5}{11}\tau^{11/3},
\end{equation} 
\noindent with
\begin{equation}
\tau=0.16\,L_{w37}^{-8/35}n_0^{-2/35}t_6^{6/35}.
\end{equation}

It is useful to express the X-ray emission in physical parameters that
are observable, such as size, expansion velocity and density. For
this, combining Eq.~\ref{radius}, \ref{veloc} and \ref{Lx}, we obtain
\begin{equation}
L_\mathrm{X}=(8.2\times 10^{27}~\mathrm{erg~s^{-1}})I(\tau)\xi R_\mathrm{pc}^{17/7}n_0^{10/7}V_5^{16/7},
\label{Lx2}
\end{equation}
\noindent where $V_5$ is the expansion velocity in units of km~s$^{-1}$.

\section{{\it XMM-Newton} observations and data reduction}

The star forming complex LMC-N\,57 was observed by {\it XMM-Newton}
with the European Photon Imaging Camera (EPIC). The observations were
performed on 2006-08-28 and correspond to the Obs.\,ID 0400320101 (PI:
R.\,Williams). The EPIC-MOS cameras were operated in the Full Frame
Mode whilst the EPIC-pn camera was operated in the Extended Full Frame
Mode. The three EPIC observations were obtained with the thin optical
filter with a total observation time of 46.71~ks. The EPIC-pn, MOS1,
and MOS2 cameras have exposure times of 42.57, 46.41, and 46.57~ks,
respectively. The {\it XMM-Newton} pipeline products were processed
using the {\it XMM-Newton} Science Analysis Software ({\sc sas}
version 16.1) and the calibration files obtained on 2019-01-15.

\begin{figure*}
\begin{center}
\includegraphics[width=\linewidth]{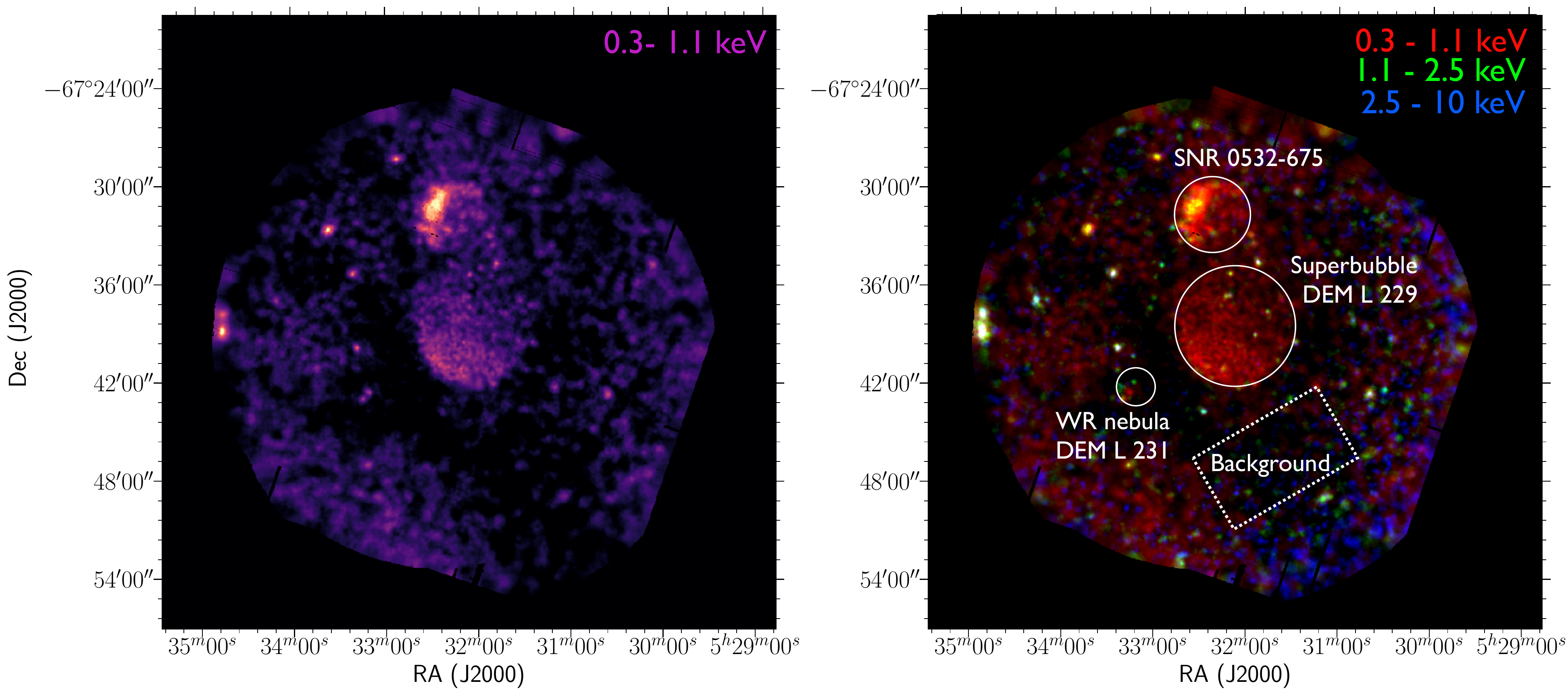}
\caption{{\it XMM-Newton} EPIC exposure-corrected,
  background-corrected images of the X-ray emission from
  LMC-N\,57. Left: Soft X-ray emission (0.3--1.1~keV). Right:
  Colour-composite image of the X-ray emission. Red, green and blue
  correspond to the soft, medium, and hard X-ray bands,
  respectively. The circular apertures show the extension of the
  diffuse X-ray emission from the SNR\,0532$-$675, the SB DEM L\,229
  and the WR nebula DEM L\,231 in LMC-N\,57. The dashed-line polygon
  shows the extraction region of the background spectrum.}
\label{fig:mosaico}
\end{center}
\end{figure*}

The Observation Data Files were processed using the {\sc sas} tasks
{\it epproc} and {\it emproc} to produce the corresponding event
files. In order to excise periods of high-background levels, we
created light curves binning data by 100~s for each of the EPIC
cameras in the 10--12~keV energy range. Background was considered high
for count rate values higher than 0.8, 0.4, and 0.4~counts~s$^{-1}$
for the pn, MOS1 and MOS2 cameras, respectively. The final useful time
for each camera was 12.67~ks (pn), 26.56~ks (MOS1), and 27.65~ks
(MOS2).

In order to produce high-quality maps of the distribution of the X-ray
emission, we used the Extended Source Analysis Software package
\citep[{\sc esas};][]{Kuntz2008,Snowden2004,Snowden2008}. The {\sc
  esas} tasks are tailored to successfully remove the contribution
from the astrophysical background, the soft proton background, and
solar wind charge-exchange reactions. A major difference between the
{\sc esas} task and those from the {\sc sas} task is that they apply
very restrictive selection criteria of events and their final net
exposure times are 7.20~ks, 10.65~ks, and 12.07~ks for the pn, MOS1,
and MOS2 cameras, respectively.

We created three EPIC images in the 0.3--1.1~keV, 1.1--2.5~keV, and
2.5--10~keV energy ranges that are labeled as soft, medium and hard
bands. Following the {\sc esas} cookbook, individual pn, MOS1, and
MOS2 images were created, merged together and, finally, corrected by
their exposure maps. The resultant exposure-map-corrected,
background-subtracted EPIC (pn$+$MOS1$+$MOS2) image of the soft X-ray
emission as well as a colour-composite image combining the three bands
are shown in Figure~\ref{fig:mosaico}. Each band image has been
adaptively smoothed using the {\sc esas} task {\it adapt} requesting
20~counts under the smoothing kernel for the soft and medium bands and
10 counts for the hard band.

\begin{figure}
\begin{center}
\includegraphics[width=\linewidth]{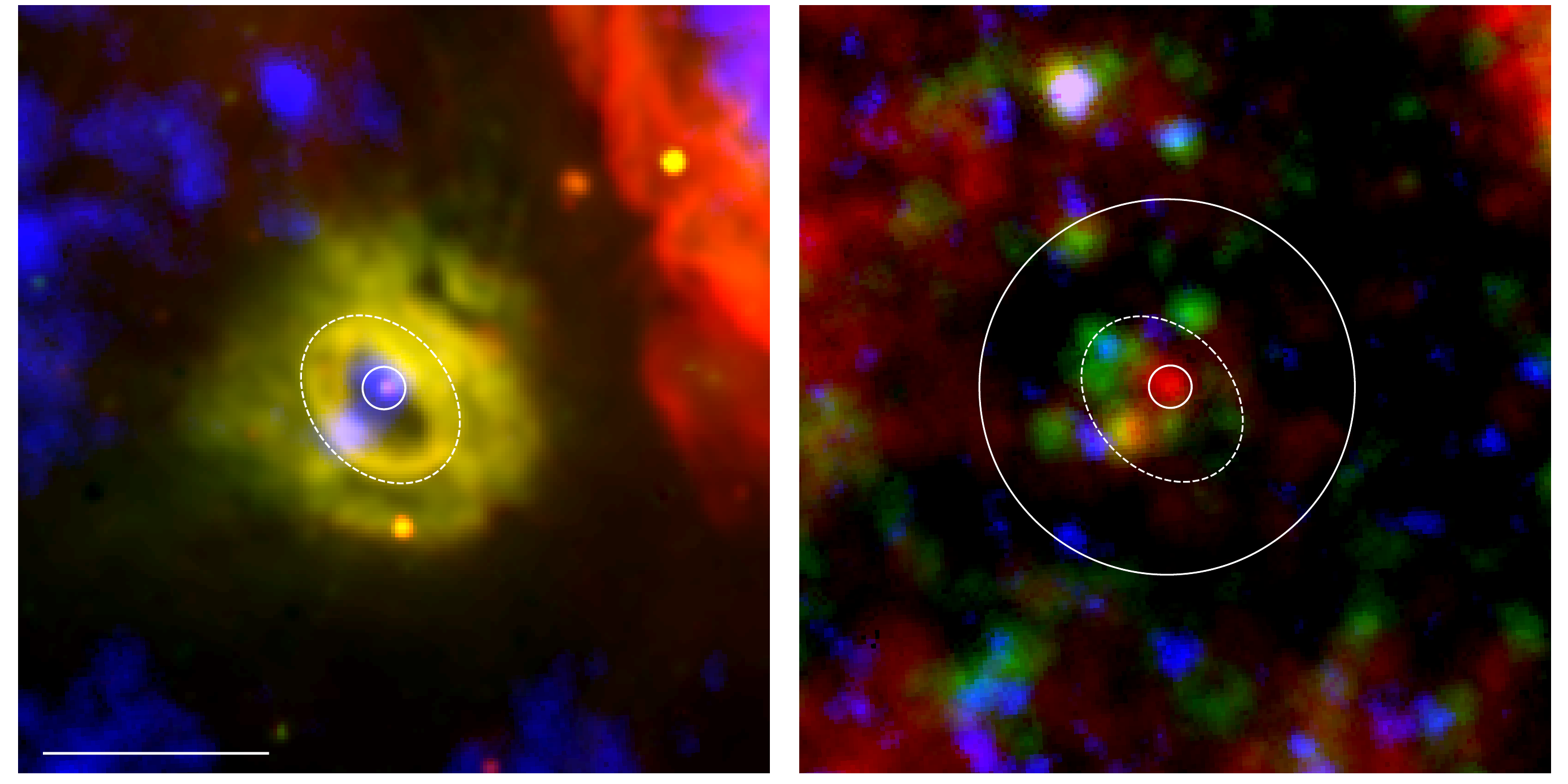}
\caption{Colour-composite images of the WR nebula DEM L\,231. Left:
  Optical and X-ray composition. Red, green and blue correspond to
  H$\alpha$, [O\,{\sc iii}] and the soft X-ray band. Right: X-ray
  composition (red - soft band, green - medium band, blue - hard
  band). The dashed-line ellipse has an extension of
  76$''\times$100$''$ while the circular aperture has a diameter of
  200$''$ and encompasses the total extension of the nebula. BR\,48 is
  the central WR star in both panels and it is marked with the
  innermost circular aperture. The white bar represents 2$'$. North is
  up and east to the left.}
\label{fig:WRnebula}
\end{center}
\end{figure}

\section{Distribution of the X-ray-emitting gas in LMC-N\,57} 

Figure~\ref{fig:mosaico} shows that the X-ray emission comes from the
three regions in LMC-N\,57. A large number of point-like sources are
also detected in the vicinity of LMC-N\,57. The SB DEM L\,229 emits
mainly in the soft X-ray band and exhibits a relatively round shape
with its maximum located towards the south-east close to the position
of the OBA LH\,76. Its angular size is 3.93$'$ in radius and
corresponds to a physical size of $\sim$60~pc. On the other hand, the
SNR\,0532$-$675 emits considerably in the medium X-ray band suggesting
noticeable spectral differences with respect of the X-ray-emitting gas
in DEM L\,229, but a similar distribution of the hot gas is
appreciated. Its maximum towards the north-east and its angular size
is $\sim$2.5$'$ with a physical size of $\sim$36~pc. No extended X-ray
emission is detected in the hard X-ray band.

\begin{figure*}
\begin{center}
\includegraphics[width=0.65\linewidth]{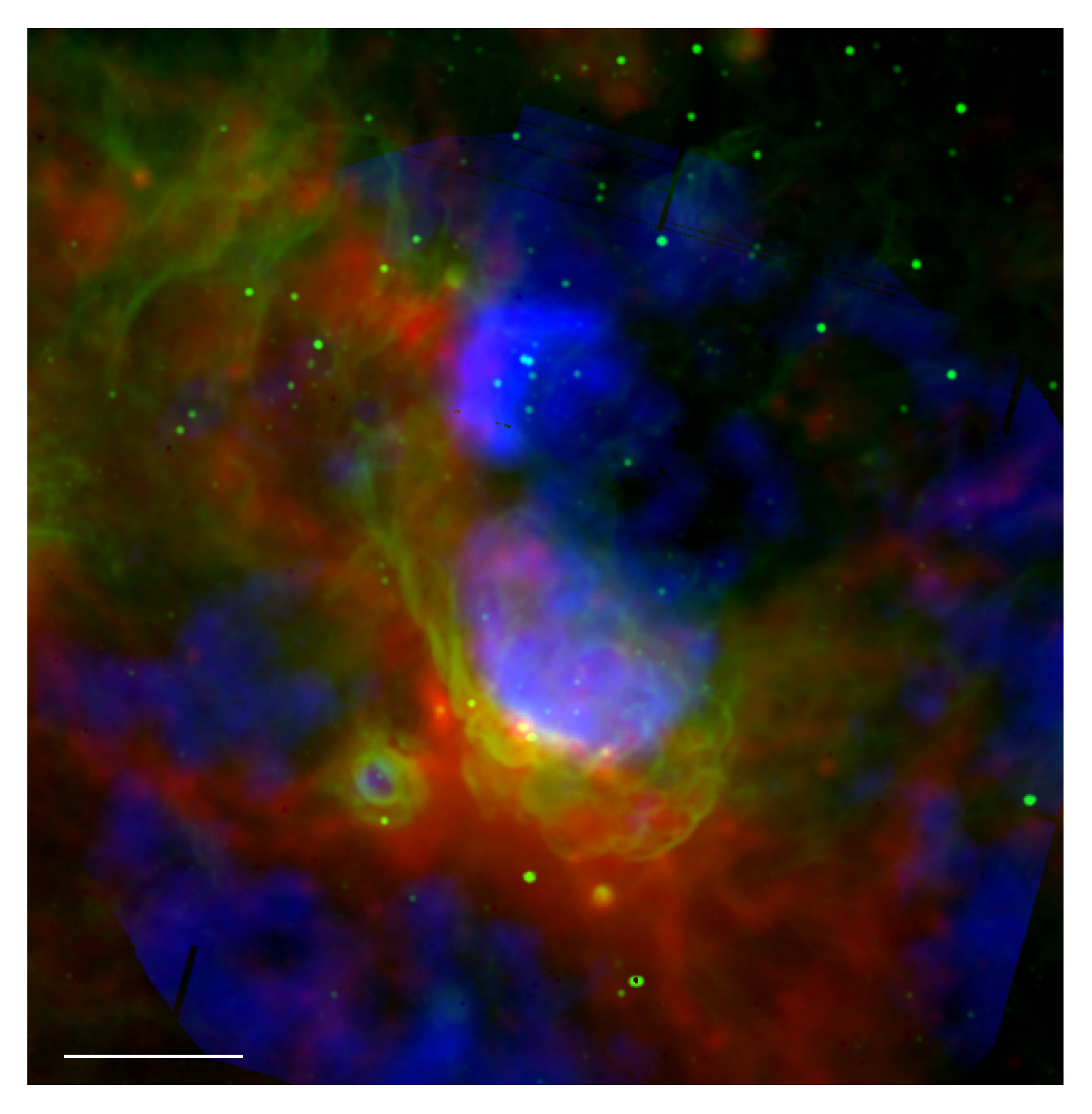}
\caption{Colour-composite image of LMC-N\,57. Red, green, and blue
  correspond to {\it Spitzer} MIPS 24~$\mu$m, H$\alpha$ from MCELS,
  and the soft X-ray band (0.3--1.1~keV). This image has the same
  field of view as those presented in Fig.~\ref{fig:rgb1}. The white
  bar represents 5$'$. North is up, east to the left.}
\label{fig:n57_rgb_final}
\end{center}
\end{figure*}

Figure~\ref{fig:mosaico} also suggests that the X-ray emission from
the WR nebula around Br\,48 is due to the presence of point
sources. This is further illustrated in
Figure~\ref{fig:WRnebula}. This figure shows that the X-ray emission
comes from (at least) 10 point sources located inside the inner shell
in DEM L\,231. We note that the presence of these sources hinders the
search and analysis of diffuse X-ray emission from DEM L\,231 as in
galactic WR nebulae \citep[see][and references therein]{toa16,toa17}.

Finally, to produce a clean view of the distribution of the diffuse
X-ray emission, we used the {\sc ciao} \citep[version
  4.9;][]{Fruscione2006} {\it dmfilth} task. All point sources were
excised from the soft X-ray band and the resultant image is compared
with the nebular H$\alpha$ and that from the {\it Spitzer} MIPS
24~$\mu$m images. Figure~\ref{fig:n57_rgb_final} shows that some faint
X-ray diffuse emission is leaking towards the north-east region, an
apparent low-density region in LMC-N\,57 (see also
Fig.~\ref{fig:rgb1}). It seems that the hot gas from the SB and from
the SNR are merging together in this low-density region. There is also
a spatial coincident between the presence of dust unveiled by the IR
observations and the lack of X-ray emission. IR emission delineates
that coming from the X-ray emission.

\begin{figure*}
\includegraphics[width=0.5\linewidth]{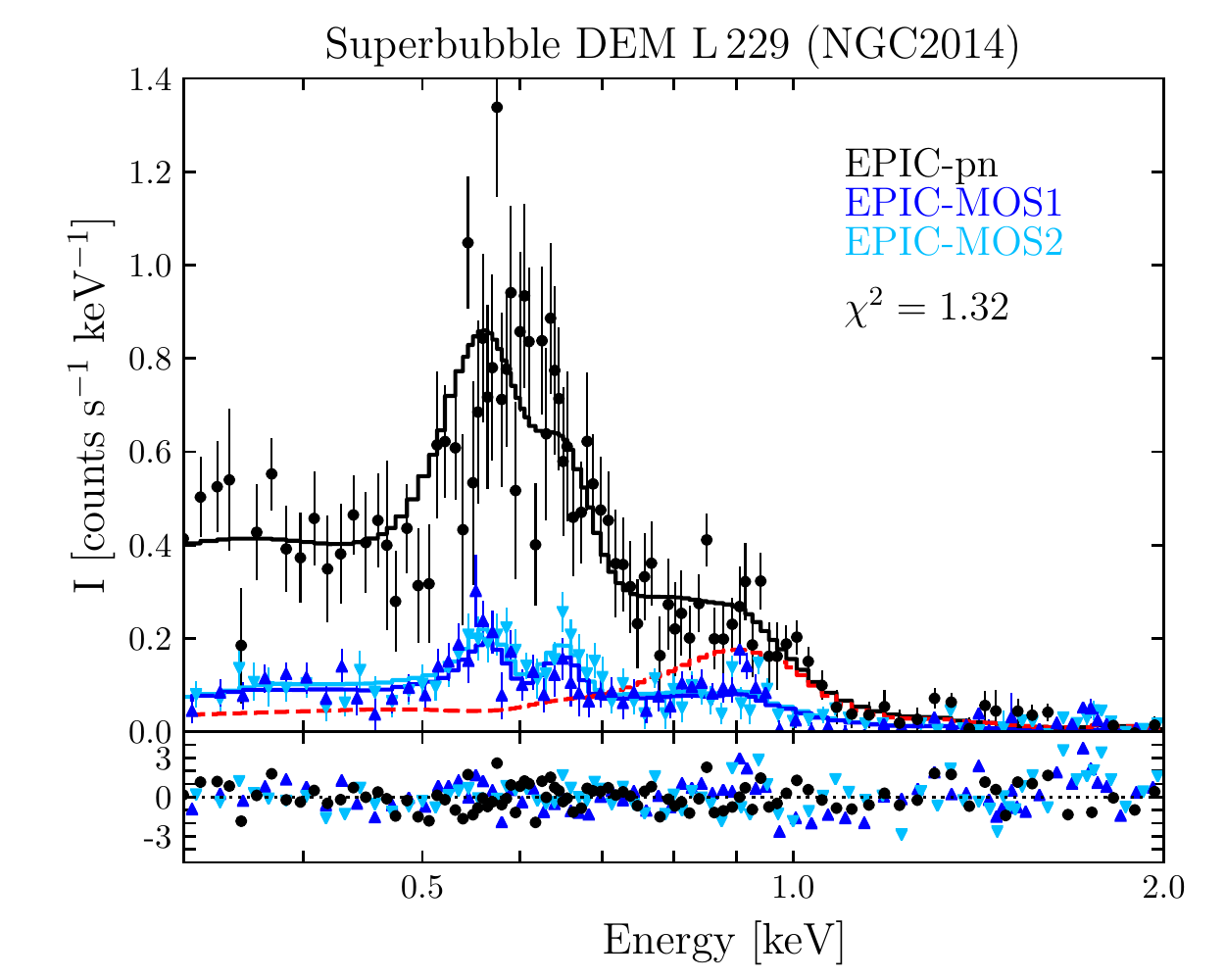}~
\includegraphics[width=0.5\linewidth]{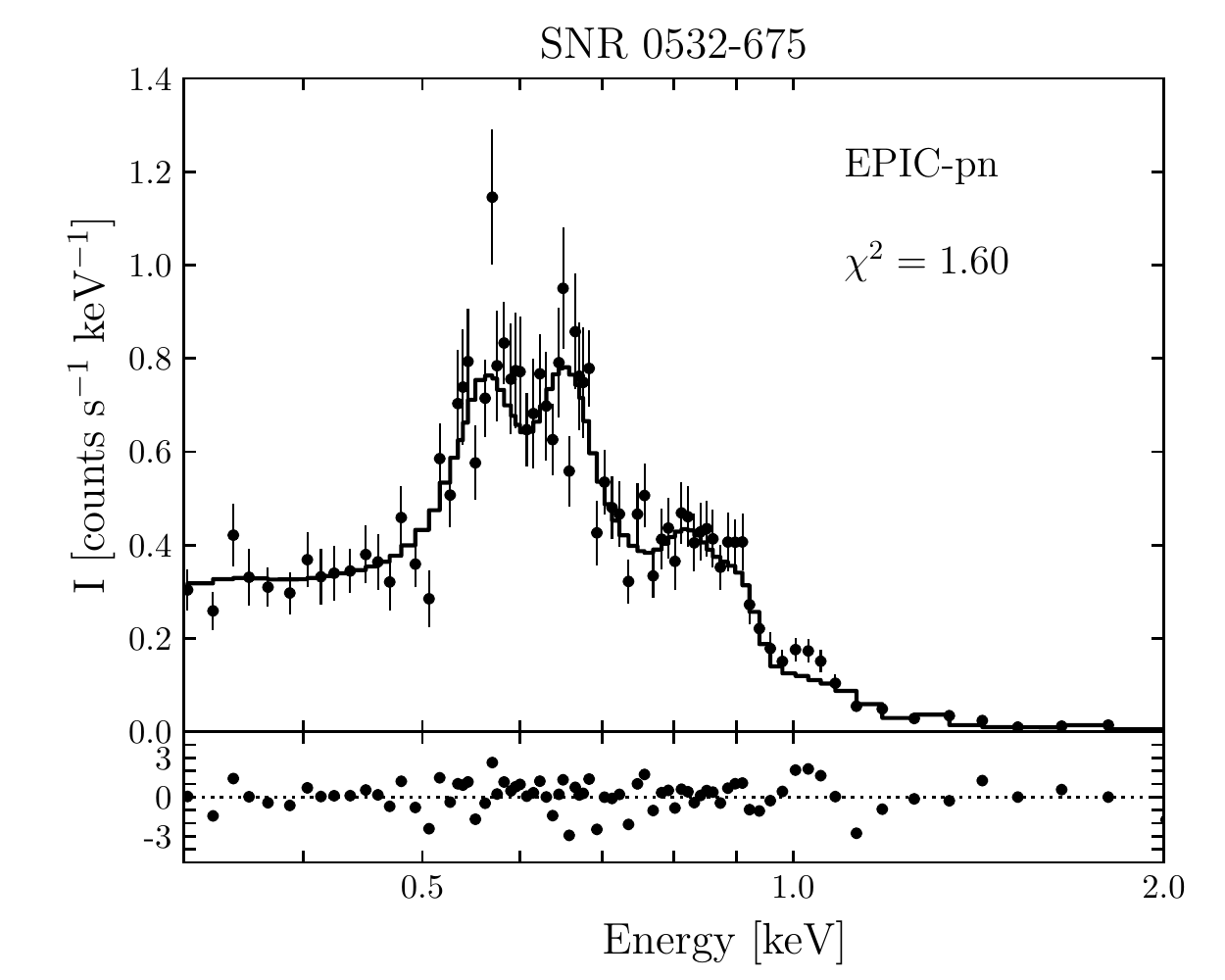}
\caption{Background-subtracted {\it XMM-Newton} EPIC spectra of the SB
  DEM L\,229 (left) and the SNR\,0532$-$675 in LMC-N\,57 (right). The
  solid lines show the best-fit model. Different symbols (colours)
  represent spectra extracted from different cameras. Residuals are
  shown in the bottom panels. The (red) dashed-line in the left panel
  represents the contribution of the higher temperature component to
  the best fit to the EPIC-pn data.}
\label{fig:spectra}
\end{figure*}

\section{Spectral analysis}

To study the physical properties of the X-ray-emitting gas we have
extracted spectra from different regions in LMC-N\,57. Circular
apertures with angular radii of 2.3$'$ and 3.7$'$ (which correspond to
33~pc and 54~pc) have been used to extract spectra from the
SNR\,0532$-$675 and DEM L\,229, respectively. The background spectrum
was extracted from a region near to DEM L\,229 with no contribution
from extended X-ray emission (see Figure \ref{fig:mosaico}). In all
cases, contaminant point-like sources have been excised. As shown in
the previous section, the presence of several point-like sources in
the vicinity of Br\,48 do not allow us to extract the spectrum of a
possible extended emission from the WR nebula DEM
L\,231. Nevertheless, we extracted the X-ray spectrum of its
progenitor WR star Br\,48. The X-ray spectra have been obtained by
using the {\sc sas} task {\it evselect} and produced the associated
calibration matrices using the {\it arfgen}, and {\it rmfgen}
tasks. The background-subtracted spectra from the SB DEM L\,229 and
the SNR\,0532$-$675 are shown in
Figure~\ref{fig:spectra}. Unfortunately, the quality of the
background-subtracted spectra of Br\,48 was low and only fits by eye
were performed (see below).

The spectral analysis was performed using {\sc xspec} \citep[version
  12.9;][]{Arnaud1996}. In accordance with previous studies of the
extended X-ray emission from SBs in the LMC we have modelled the X-ray
spectra using the absorbed {\it vapec} optically-thin plasma emission
model with {\it tabs} absorption model
\citep[][]{Wilms2000}. Initially, all abundances were set to ISM
values of the LMC, but in order to improve the statistics of the fit,
we left some elements as free parameters. Table \ref{tab:abund} lists
the abundance values initially adopted.

The resultant model spectra were compared with the observed X-ray
spectra in the 0.3--5~keV energy range where $\chi^{2}$ statistics was
used to evaluate the goodness of the fits. A minimum of 50 counts per
bin was requested for the spectral fits.

\begin{table}[t]
\begin{center}
\caption{Adopted LMC abundances used for spectral fit}
\begin{tabular}{cccl}
\hline 
\hline
Element   & $X/X_{\odot}$   & (log$_{10}X+ 12$) & Reference\\
\hline
C         & 0.20            & 7.90         & \citet{Korn2000}\\
N         & 0.47            & 7.72         & \citet{Korn2000}\\
O         & 0.21            & 8.25         & \citet{Maggi2016}\\
Ne        & 0.28            & 7.53         & \citet{Maggi2016}\\
Mg        & 0.33            & 7.10         & \citet{Maggi2016}\\
Si        & 0.69            & 7.39         & \citet{Maggi2016}\\
S         & 0.36            & 6.76         & \citet{Schenck2016}\\
Fe        & 0.35            & 7.21         & \citet{Maggi2016}\\
\hline
\hline 
\end{tabular}
\end{center}
\label{tab:abund}
\end{table}

\subsection{The superbubble DEM L\,229}

The EPIC spectra of DEM L\,229 are shown in Fig~\ref{fig:spectra} left
panel. The SB was registered by the three EPIC cameras. The most
important spectral features are a dominant X-ray peak between
0.5--0.6~keV which might be attributed to the O\,{\sc vii} at 0.58~keV
and a secondary peak more clearly seen in the MOS spectra appears
between 0.6--0.7~keV and can be attributed to the O\,{\sc viii} at
0.65~keV. The total count rate of the pn, MOS1, and MOS2 spectra are
330~counts~ks$^{-1}$, 77.5~counts~ks$^{-1}$, and 96~counts~ks$^{-1}$
and correspond to a total photon counts of 4180~counts, 2060~counts,
and 2660~counts, respectively.

According to \citet{Hainich2014} the WR star BAT99 56 is located at
the edge of the SB DEM L\,229. Its estimated $E(B-V)$ is 0.12 and
translates into a column density of
$N_\mathrm{H}$=6.96$\times10^{20}$~cm$^{-2}$. Thus, for this region we
fixed this value for the following spectral analysis.

We first fitted the EPIC-pn spectrum with a two-temperature plasma
emission model. The best-fit model ($\chi^{2}=1.05$) resulted in
plasma temperatures of $kT_{1}=0.19^{+0.1}_{-0.01}$~keV and
$kT_{2}=0.94^{+0.10}_{-0.09}$~keV with normalization parameters
$A_{1}$ and $A_{2}$ of $9.7\times10^{-4}$~cm$^{-4}$ and
$1.9\times10^{-4}$~cm$^{-4}$, respectively. Most of the element
abundances converged to their ISM parameters except for the oxygen
abundance. The best model suggest an overabundance of oxygen of
0.48$^{+0.14}_{-0.10}$ times solar, twice the initial ISM oxygen
abundance (see Table~1). The absorbed flux in the 0.3--5.0~keV was
$f_\mathrm{X}=(6.9\pm0.2)\times10^{-13}$~erg~cm$^{-2}$~s$^{-1}$, which
corresponds to an intrinsic flux of
$F_\mathrm{X}=(1.2\pm0.1)\times10^{-12}$~erg~cm$^{-2}$~s$^{-1}$. Thus,
we estimated an X-ray luminosity of
$L_\mathrm{X}=(3.5\pm0.3)\times10^{35}$~erg~s$^{-1}$.

Subsequently, we fitted simultaneously the three EPIC spectra. The
best-fit model ($\chi^{2}$=1.32) resulted in a two-temperature plasma
emission model with temperatures $kT_{1}=0.19^{+0.01}_{-0.06}$~keV and
$kT_{2}=0.86^{+0.05}_{-0.06}$~keV. The resultant normalization
parameters for each component are $A_{1}=9.4\times10^{-4}$~cm$^{-5}$
and $A_{2}=1.9\times10^{-4}$~cm$^{-5}$, respectively. This model was
achieved by leaving the oxygen abundance as a free parameter. This
converged to 0.52$^{+0.12}_{-0.08}$ times its solar value
\citep{Anders1989}. The absorbed and intrinsic fluxes in the
0.3--5.0~keV energy range are very similar to the fit performed to the
EPIC-pn spectrum (see above):
$f_\mathrm{X}=(6.9\pm0.2)\times10^{-13}$~erg~cm$^{-2}$~s$^{-1}$ and
$F_\mathrm{X}=(1.2\pm0.1)\times10^{-12}$~erg~cm$^{-2}$~s$^{-1}$. The
estimated X-ray luminosity is
$L_\mathrm{X}=(3.5\pm0.3)\times10^{35}$~erg~s$^{-1}$. This model is
plotted along side the background-subtracted EPIC spectra in Fig.~6
left panel. The contribution from the second temperature component to
the EPIC-pn spectrum is also shown with a dashed (red) line.

Finally, we note that more complicated models, adding a power law
component were also attempt in order to fit the marginal emission at
energies above 2.0~keV, but we note that the best fit was not
statistically better. For example, in the simultaneous fit to the
three EPIC, the model was not as good ($\chi^{2}>1.43$) as the one
described above. We further discuss this in Appendix~A. Another model
adopting a non-equilibrium ionisation model (see next subsection) was
also attempted but resulted in unrealistic column density values and
extremely high plasma temperatures.

\subsection{SNR 0532$-$675}

The SNR\,0532$-$672 was registered by the EPIC-pn and MOS2
detectors. Unfortunately, CCD\,6 was no longer functional in the MOS1
camera by the time the observations were performed and corresponds to
the area where the SNR should have been registered \citep[see
  appendix~F in][]{Maggi2016}. Thus, we only show the EPIC spectra
extracted from the pn and MOS2 cameras in Fig.\ref{fig:spectra} right
panel. Similarly to the SB DEM L\,231, the most prominent lines are
those of the O\,{\sc vii} at 0.58 and O\,{\sc viii} at 0.65~keV,
although the later is brighter in these spectra. The resultant count
rate for the pn and MOS2 spectra are 330~counts~ks$^{-1}$ and
95~counts~ks$^{-1}$ which correspond to a total of 4180 and
2630~counts, respectively.

Similarly to the SB DEM L\,299, we first modelled the EPIC-pn
spectrum. The best-fit model ($\chi^{2}=1.20$) with plasma
temperatures components of $kT_{1}=0.20^{+0.10}_{-0.10}$ and
$kT_{2}=0.72^{+0.04}_{-0.04}$ with normalization parameters of
$A_{1}=6.8\times10^{-4}$~cm$^{-5}$ and
$A_{2}=3.5\times10^{-4}$~cm$^{-5}$. The oxygen abundance turned out be
0.82$^{+0.22}_{-0.15}$ times its solar value. The absorbed and
intrinsic fluxes in the 0.3--5.0~keV are
$f_\mathrm{X}=(9.3\pm0.2)\times10^{-13}$~erg~cm$^{-2}$~s$^{-1}$ and
$F_\mathrm{X}=(1.5\pm0.1)\times10^{-12}$~erg~cm$^{-2}$~s$^{-1}$. The
corresponding X-ray luminosity is
$L_\mathrm{X}=(4.5\pm0.2)\times10^{35}$~erg~s$^{-1}$.

However, we note that a simpler model can be achieved by adopting a
non-equilibrium ionisation (NEI) model as that described by
\citet{Maggi2016}. We modelled the EPIC-pn spectrum of SNR\,0532$-$675
with a {\it vpshock}, adopting those abundaces found by
\citet{Maggi2016} and found very similar results as those listed in
their table~E.1. We fixed the plasma temperature to that obtained by
those authors ($kT$=0.53~keV) and the best-fit model ($\chi^{2}=1.60$)
resulted in a column density
$N_\mathrm{H}=(9.6\pm0.2)\times10^{20}$~cm$^{-2}$ and ionisation
time-scale $\tau=(10.3\pm2.2)\times10^{10}$~s~cm$^{-3}$. This model
yielded an absorbed X-ray flux of
$f_\mathrm{X}=(8.8\pm2.4)\times10^{-13}$~erg~cm$^{-2}$~s$^{-1}$ with
an intrinsic flux of
$F_\mathrm{X}=(1.7\pm0.5)\times10^{-12}$~erg~cm$^{-2}$~s$^{-1}$, which
corresponds to an X-ray luminosity of
$L_\mathrm{X}=(5.0\pm1.0)\times10^{35}$~erg~s$^{-1}$. This model is
presented in Figure~6 in comparison with the EPIC-pn spectrum.

We remark that no further discussion on the X-ray properties of
SNR\,0532$-$675 will be pursuit in this work as it has been previously
done by \citet{Maggi2016}.

\subsection{The WR star Br\,48}

As mentioned before, there are several point-like sources projected
inside DEM L\,231 hampering a proper search and analysis of extended
emission in this WR nebula. Thus, we only concentrate in the analysis
of the X-ray emission from its progenitor WR star, Br\,48.

DEM L\,231 was registered by the three EPIC cameras. Unfortunately,
its resultant count rates are low and a detailed spectral analysis is
not possible. The EPIC-pn, MOS1, and MOS2 count rates are
0.63~counts~ks$^{-1}$, 0.37~counts~ks$^{-1}$, and
0.18~counts~ks$^{-1}$ with total photon counting of 8, 10, and 5
counts, respectively.

The column density was fixed to
$N_\mathrm{H}$=9.3$\times10^{20}$~cm$^{-2}$ estimated from the
$E(B-V)$ value of 0.16 reported by \citet{Hainich2014}. The observed
spectrum can be reasonably well described by a plasma emission model
with temperature $kT$=0.40~keV. We estimated the absorbed and
intrinsic fluxes as
$f_\mathrm{X}$=1.5$\times10^{-15}$~erg~cm$^{-2}$~s$^{-1}$ and
$F_\mathrm{X}$=2.5$\times10^{-15}$~erg~cm$^{-2}$~s$^{-1}$,
respectively. The estimated X-ray luminosity is
$L_\mathrm{X}$=7.3$\times10^{32}$~erg~s$^{-1}$ which is consistent to
what is found for other WR stars in the LMC
\citep[][]{GuerreroChu2008} and in our Galaxy
\citep[][]{Oskinova2015}.

\section{Discussion}

\begin{table}
\begin{center}
\caption{Spectral types, terminal velocities and mass loss rates for the stars of the SB DEM L\,229$^a$}
\begin{tabular}{c c c c}
\hline
Star   &    Spectral type    & $V_{\infty}$ (km s$^{-1}$) &  log$_{10}(\dot M)$  \\
\hline \hline
L1     &      O4If           &  2500                      &  -4.92          \\
L21    &      O4If           &  2500                      &  -4.92          \\
L51    &      B0Ia           &  1800                      &  -5.15          \\
L59    &      B2Iab          &  1500                      &  -5.37          \\
L70    &      B0III          &  1600                      &  -6.90          \\
S4     &      O8V            &  1900                      &  -6.89          \\
S6     &      O8I            &  2000                      &  -5.04          \\
\hline \hline
\end{tabular}
\end{center}
\center$^a$The stellar content of LH\,74 was taken from \citet{w96} whilst their mass-loss rates were obtained from \citet{dejag88}.
\label{tab:tspec}
\end{table}

\citet{dunne01} presented the first analysis of X-ray observations of
LMC-N\,57 along with another 12 star forming regions in the LMC. Their
{\it ROSAT} X-ray observations were able to unambiguously detect
X-rays from the SNR\,0532$-$675 \citep[previously discovered by][using
  radio and X-ray observations from MOST and {\it Einstein}
  telescopes]{mat85}. Although LMC-N\,57 was detected near the edge of
the {\it ROSAT} PSPC detectors, where the point-spread function is
bad, the authors found that the X-ray-emitting gas from the DEM L\,229
was confined by the nebular material \citep[see
  also][]{Points2001}. No X-ray emission was detected from the WR
nebula DEM L\,231.

The analysis of the {\it XMM-Newton} observations of LMC-N\,57
presented here have unveiled, in unprecedented detail, the
distribution of its X-ray emission. The X-ray-emitting gas within DEM
L\,229 is nicely delineated by its nebular emission detected in
H$\alpha$ and the dust emission detected with the {\it Spitzer} MIPS
24~$\mu$m toward the south, but it seems to be leaking out to a low
density region toward the north and, very likely, combining with the
extended emission from the SNR\,0532$-$675 (see
Fig.~\ref{fig:mosaico}). It is clear that the SB extends further
toward the south but not diffuse X-ray emission is detected in this
region. It might be due to the higher column density. Inspection of
archival {\it Herschel} PACS and SPIRE observations (not shown here)
suggest at the presence of very cold dust toward the southern region
of DEM L\,229.

We can theoretically estimate the X-ray luminosity from the SB DEM
L\,229 by using the equations described in Section~2. First, we
searched for the massive stellar content in DEM L\,229 presented in
\cite{w96}. Table~\ref{tab:tspec} lists the spectral types and
terminal wind velocities of the most massive stars in LH\,76. We have
included the characteristic values for the mass-loss rates associated
with stars of such spectral types from \citet[][]{dejag88}. In order
to use Eq.~5 we need to estimate the ISM density, which was obtained
from Eq.~\ref{dens} and a dynamical age of 0.8 Myr was also considered
(Eq.~\ref{radius}). The estimated X-ray luminosity is
$L_\mathrm{X}$=3.1$\times$10$^{35}$~erg~s$^{-1}$. If we now use Eq.~8
to estimate the X-ray luminosity, adopting a physical radius of 60~pc
and an expansion velocity for DEM L\,229 of 45~km~s$^{-1}$
\citep{Chu1999} we obtain
$L_\mathrm{X}$=1.6$\times$10$^{35}$~erg~s$^{-1}$. Both estimates of
the X-ray emission from DEM L\,229 are very similar to the estimated
value from the observations (see Section~5.1), also very similar to
that estimated by analytical calculations presented in \citet{dunne01}
(see column~5 in table~5 of that paper). We can only conclude that the
X-ray emission from the SB DEM L\,229 is the product of the
combination of the stellar winds from the massive stars in LH\,74. In
accordance to \citet{mc} (see Section~1), we found that there is no
need to suggest that a SN has exploded inside DEM L\,229 to enhance
the X-ray emission.

We note, that the X-ray luminosity presented in \citet{dunne01} and
obtained from spectral fitting, that is
8.1$\times$10$^{35}$~erg~s$^{-1}$, is twice the luminosity obtained in
Secion~5.1. This discrepancy can be amended by considering the limited
spatial resolution of {\it ROSAT} satellite which made it difficult to
identify and excise unrelated point-like sources in the field of view
of DEM L\,229. The present analysis of {\it XMM-Newton} observations
of LMC-N\,57 call for a detailed analysis of the distribution of the
X-ray-emitting gas and spectral properties from star forming complexes
in the MCs.

It is interesting to notice that most star forming regions in the LMC
that have been studied in X-rays have relatively high X-ray
luminosities compared to LMC-N\,57 \citep[see,
  e.g.,][]{dunne01,Jaskot2011,reyes14,zhang14}. The fact that these
high X-ray luminosities can not be explained purely by the
pressure-driven bubble model (see Section~2) has led to the suggestion
that the contribution from SN explosions power the X-ray emission to
such high observed luminosities ($\lesssim10^{37}$~erg~s$^{-1}$). In
particular, the SN explosions that occur very close to the edges of
the complex hitting the nebular material \citep[][]{chum90}. The
latter also corroborated by numerical simulations
\citep[e.g.,][]{ary11}.

Unfortunately, the search of extended X-ray emission from the WR
nebula DEM L\,231 is hampered by the presence of other point-like
sources projected inside the nebula. We note that DEM L\,231 is also
surrounded by a dust-rich medium which might easily extinguish the
soft X-ray emission. Nevertheless, we were able to extract the X-ray
spectrum of Br\,48, the so-called progenitor star of DEM
L\,231. \citet{Hainich2014} presented the analysis of Br\,48 by means
of the stellar atmospheres code
PoWR\footnote{\url{http://www.astro.physik.uni-potsdam.de/~wrh/PoWR/powrgrid1.php}}
and concluded that this WR stars is a binary of the WN4$+$O9 type. We
note, however, that the same group have produced a more recent
analysis of WR binaries in the LMC. Their most recent analysis of
Br\,48 (a.k.a., BAT99 59) suggests that this WR star is composed by a
WN3$+$O6\,III system \citep[see][]{Shenar2019}.  The bolometric
luminosity of their best-fit model is
log$_{10}(L/\mathrm{L}_{\odot})$=6.45. This means that Br\,48 fulfills
the $L_\mathrm{X}/L_\mathrm{bol}\sim-7$ relation exhibit in O-star
binaries and WN binary stars \citep[see figure 1 in][]{Oskinova2015}.

The three component regions of LMC-N\,57 seem to have been formed from
the same giant filamentary structure unveiled by IR observations (see
Fig.~1). If one assumes that the progenitor stars of these three
component regions were formed at the same time, we can argue that the
most massive stars were located towards the north of LMC-N\,57, where
the SNR\,0532-675 was formed. The next massive star would be the WR
star Br\,48, which is about to explode as a SN inside DEM L\,231. We
have demonstrated that no SN explosion has occurred inside DEM L\,229,
that is, the less massive hot stars are those from the LH\,74 star
cluster. LMC-N\,57 is a clear example of the effect of massive stars
in their environment. In a radius of $\sim$150~pc massive stars in
different stages of evolution destroying their natal cloud.

The fact that the X-ray luminosity of the SB can be explained by only
accounting for the contribution from the stellar winds of the hot
stars in DEM L\,229 led us to suggest that the oxygen enhancement
might be due to mixing with the X-ray-emitting gas from the SNR as
hinted by Figure~5. Nevertheless, we note that the oxygen abundance of
the SB DEM L\,229 is only twice the averaged value reported by
\citet{Maggi2016}. These authors list oxygen abundances within 0.1 and
0.5 times the solar value in their figure~5, that is, the oxygen
abundance is still somewhat consistent with ISM measurements in the
LMC.

Finally, we note that the estimated X-ray luminosity of the
SNR\,0532$-$675 obtained here is consistent with theoretical
calculations. \citet{leahy17} estimated an age of $\sim 22$ kyr, which
is a typical value of the Sedov-Taylor phase in SNRs. Using equation~1
from \citet{chum90}, adopting a radius of $r=36$~pc, an abundance of
0.3 times solar and an ISM density of $n_{0}\sim 0.08$~cm$^{-3}$
\citep{leahy17}, we can estimate an X-ray luminosity of
$L_\mathrm{X}=4.4\times10^{35}$~erg~s$^{-1}$. In close agreement of
what was reported in Section~5.2.

\subsection{On the pressure components of DEM L\,229}
 
We have calculated the direct radiation pressure $P_\mathrm{dir}$
assuming a spherical H\,II region and eq. 3 in \citet{pel} used in the
following form
\begin{equation}
    P_\mathrm{dir}=\frac{Q_o<h\nu>}{4\pi R^2c}.
\end{equation}
\noindent Here, $Q_o=10^{50}$~photon~s$^{-1}$ is the ionizing photon
flux emitted by the star cluster, $R=60$ pc is the radius of the
SB, $<h\nu>=15$~eV is the mean photon energy by an O-type star, and
$c$ is the speed of light. This gives
$P_\mathrm{dir}=0.018\times10^{-11}$~dyn~cm$^{-2}$

The electron density of the optical shell $n_\mathrm{e}$ can be
obtained through the {\it{rms}} electron density $n_\mathrm{e,rms}$
determined from the H$\alpha$ surface brightness, by using the
equations 8-10 from \citet{chu95}. In the case of DEM L\,229, an
emission measure of $EM$=1500~cm$^{-6}$~pc \citep{dunne01} and an
electron density of $n_\mathrm{e}$=8.7~cm$^{-3}$ were obtained. Thus,
the pressure of the nebular material is estimated to be
$P_\mathrm{HII}=1.20\times10^{-11}$~dyn~cm$^{-2}$ (adopting a
temperature of 10$^{4}$~K).

To estimate the pressure of the X-ray-emitting gas, we need to compute
its electron density. For this, we can use the definition of the
normalization parameter $A$ obtained from our spectral fitting in
Section~5. This parameter can be written in a simplified form as
\begin{equation}
A= 10^{-14}\times\frac{\int{n_\mathrm{e,X}^{2} dV}}{4\pi d^2},
\end{equation}
\noindent where $V$ is the volume of the X-ray-emitting region and $d$
is the distance to the source.

We first assume a spherical morphology and, using the normalization
parameter of the best-fit model to the diffuse X-ray emission of DEM
L\,229, we obtain an electron density of
$n_\mathrm{e,X}$=0.04~cm$^{-3}$. Thus, the pressure of the
X-ray-emitting gas in
$P_\mathrm{X}$=1.22$\times$10$^{-11}$~dyn~cm$^{-2}$. Although this
value is of the order of that estimated for the ionized material, it
only represents a lower limit value. According to the wind-blown
bubble model described by \citet{w77}, which describes fairly well the
X-ray emission from DEM L\,229, the soft X-ray emission comes only
from a conduction layer at the outer edge of the hot bubble. We note
that even in 2D and 3D numerical simulations of the formation of hot
bubbles around single and groups of massive stars, this conductive
layer can become unstable and form clumps and filaments close to the
outer edge of the hot bubble. Still, the idea remains and the soft
X-ray material has a relatively small volume compared to the total
volume of the hot (unmixed) bubble \citep[e.g.,][]{dw13}.

If we take a conservative value for the thickness of this mixing
region, for example, 10 per cent of the total radius of the hot bubble
the electron density is $n_\mathrm{e,X}$=0.065~cm$^{-3}$ and the
X-ray-emitting gas pressure is
$P_\mathrm{X}$=2$\times$10$^{-11}$~dyn~cm$^{-2}$. Here we neglect the
contribution of direct radiation pressure $P_\mathrm{dir}$, because
this one is two orders of magnitude less than $P_\mathrm{HII}$ or
$P_\mathrm{X}$. Thus, the dynamics and evolution of the SB DEM L\,229
is dominated by the pressure of its hot gas powered by the combination
of stellar winds from the cluster of stars LH\,76.

Assuming that the diffuse X-ray emission uniformingly fills the
observed bubbles, is not an accurate way to estimate the density of
the X-ray-emitting material nor its pressure. We note that both
regions in the hot bubble, the unmixed material and that suffering
from the thermal conductivity effect, have the same pressure
\citep[see for example figure~5 in][]{toa11}. Thus, the unmixed region
of the hot bubble has a larger temperature, but a lower electron
density. So that pressure remains constant through the hot bubble. We
argue that the procedure used by \citet{lopez} is not correct and
underestimated the pressure values of their sample. Furthermore, we
have also demonstrated here the effective area and angular resolution
of the {\it XMM-Newton} compared to {\it ROSAT}. An improved analysis
of the diffuse X-ray emission from the superbubbles in the MCs should
be performed with higher-quality X-ray observations in order to shed
light into the discussion about the pressure role of hot gas in star
forming regions.

\section{Conclusions}

We have analyzed the {\it XMM-Newton} X-ray observations of the star
forming region LMC-N\,57. This complex is composed by a central SB DEM
L\,229, the SNR\,0532$-$675, and the WR nebula DEM L\,231 around the
WR star Br 48. Our observations allowed us to unveil the distribution
of the X-ray-emitting gas with unprecedented detail. The comparison
with optical and IR observations suggest that the combination of the
feedback from the massive stars in these three regions are destroying
their natal molecular cloud. Our findings can be summarized as:
\begin{itemize}
    \item The analysis of the distribution of the X-ray-emitting gas
      in LMC-N\,57 confirms that DEM L\,229 is filled with hot
      gas. The hot bubble is delimited by the southern molecular
      material but seems to be leaking out towards the northern
      region. The hot gas seems to be leaking out toward a low-density
      region in the northern region and, very likely, mixing with the
      hot gas produced by the SNR\,0532$-$675.
    \item We found that the X-ray emission from the SB DEM L\,229 is
      consistent with pressure-driven model. The X-ray luminosity
      estimated taking into account the stellar winds from the hot
      stars within this SB
      ($L_\mathrm{X}=1-3\times10^{35}$~erg~s$^{-1}$) is very similar
      to that obtained from the spectral analysis of the {\it
        XMM-Newton} EPIC observations,
      $L_\mathrm{X}=(3.5\pm0.3)\times10^{35}$~erg~s$^{-1}$. There is
      no need to invoke the idea that a SN has exploded within DEM
      L\,229.  We argue that this finding is supported by an improved
      analysis of the X-ray observations compared to previous {\it
        ROSAT} observations. The current EPIC observations allow a
      better identification of contaminant point-like sources within
      DEM L\,229.
    \item The pressure-driven model used here suggests that the
      X-ray-emitting gas is produced at a conduction layer relatively
      thin compared to the size of the hot bubble. Adopting a
      conservative thickness for this conduction layer we estimate
      that the pressure of the hot gas is larger than the pressure of
      the ionized material. We argue that such geometric consideration
      should be taken into account when computing pressure values of
      for the X-ray-emitting gas from star forming regions. Finally we
      note that $P_\mathrm{dir}<P_\mathrm{HII}<P_\mathrm{X}$ means
      that the radiation pressure does not contribute to the expansion
      of DEM L\,229.
    \item No diffuse X-ray emission is detected from the WR nebula DEM
      L\,231. IR observations show the presence of cold dust
      surrounding DEM L\,231, suggesting that the presence of soft
      extended X-ray emission might be easily extinguished by the high
      column density.
    \item X-ray emission is detected from the WR star Br\,48. Its
      luminosity is found to be
      $L_\mathrm{X}$=7.3$\times$10$^{32}$~erg~s$^{-1}$ which is
      consistent with the empiric $L_\mathrm{X}/L_\mathrm{bol}\sim-7$
      ratio. This is in line with the suggestion that Br\,48 is a
      WN3$+$O found by previous works.
    
\end{itemize}

We remark that {\it XMM-Newton} EPIC is currently the best instrument
to produce high-quality studies of star forming regions in the
MCs. Its large field of view and the unrivaled effective area in the
soft X-ray band makes it a perfect instrument to produce clean view of
the distribution and spectral properties of hot bubbles as well as
producing pressure estimates. The present work is the first of a
series of papers in which we revise the role feedback from massive
stellar clusters from the MCs using {\it XMM-Newton}.

\section*{Acknowledgements}

IR-B and JAT are funded by UNAM DGAPA PAPIIT project IA100318. JR-I
acknowledges finantial support from PRODEP-SEP, TNM-SEP, DGAPA-PAPIIT
(UNAM) grant IG100516. MR thanks finantial support from DGAPA-PAPIIT
(UNAM) IN109919 and CONACYT CY-253085 grants.

\appendix

\section{Searching for non-thermal X-ray emission}

SBs are the result of the contribution of stellar winds and SN
explosions. SN activity is thought to contribute to the X-ray
properties of the SB with thermal and non-thermal emission. The latter
could be unveiled by the spectral signature of a power-law
contribution to the spectral fit \citep[e.g.,][]{bam04}, but this has
been proven to be a difficult task. For example, \citet{coop04} argued
that the SB DEM L\,192 (LMC-N51) required a power-law with a photon
index of $\Gamma=1.3$ to fit its EPIC X-ray spectrum for energies
above 1~keV whilst the emission line hot plasma dominates the X-ray
emission below this energy. Similarly, \citet{Maddox2009} found that
the best-fit to the X-ray spectrum of LMC-N\,11 could be achieved by a
power-law up to energies above 5~keV. Nevertheless, \citet{yam10}
presented a detailed study of N\,11 and N\,51 combining {\it Suzaku}
and {\it XMM-Newton} data and found that the claims of presence of a
non-thermal component are associated to inaccurate background
subtraction.

We have shown in Section~5 that a second plasma temperature is
required to achieve a good fit to the X-ray spectra of the SB DEM
L\,229. This second plasma component is used to fit the emission
$\gtrsim$1~keV. Furthermore, we show in Fig.~7 the EPIC-pn spectra of
the SB presented in Fig.~6 (left panel) but in logarithmic scale. The
best-fit two-temperature plasma emission models to the X-ray emission
of the SB fairly fits the emission in the high energy range (shown in
solid black line in Fig.~7). The contribution from the second
component (shown in red dashed-line in Fig.~7) mainly fits the high
energy part of the spectrum. In order to assess the presence of
possible non-thermal emission from the SB DEM L\,229, we have further
included a power-law component to the spectral analysis presented in
Section~5. Such complex model did not improved the model described in
Section~5, thus, it was discarded. On the other hand, the a one
temperature NEI plasma model that best fits the EPIC spectrum of the
SNR\,0532$-$675 can model the background subtracted spectrum in the
complete range (see Fig.~7 right panel). A more complicated model,
adding a power-law component to the fit did not improved the model
described in Section~5. Hence, within the quality of the present EPIC
spectra we can discard the idea of non-thermal X-ray emission from
these regions. Although we note that the spectra quality in the high
energy range is not good (see Fig.~7).

\begin{figure*}
\includegraphics[width=0.5\linewidth]{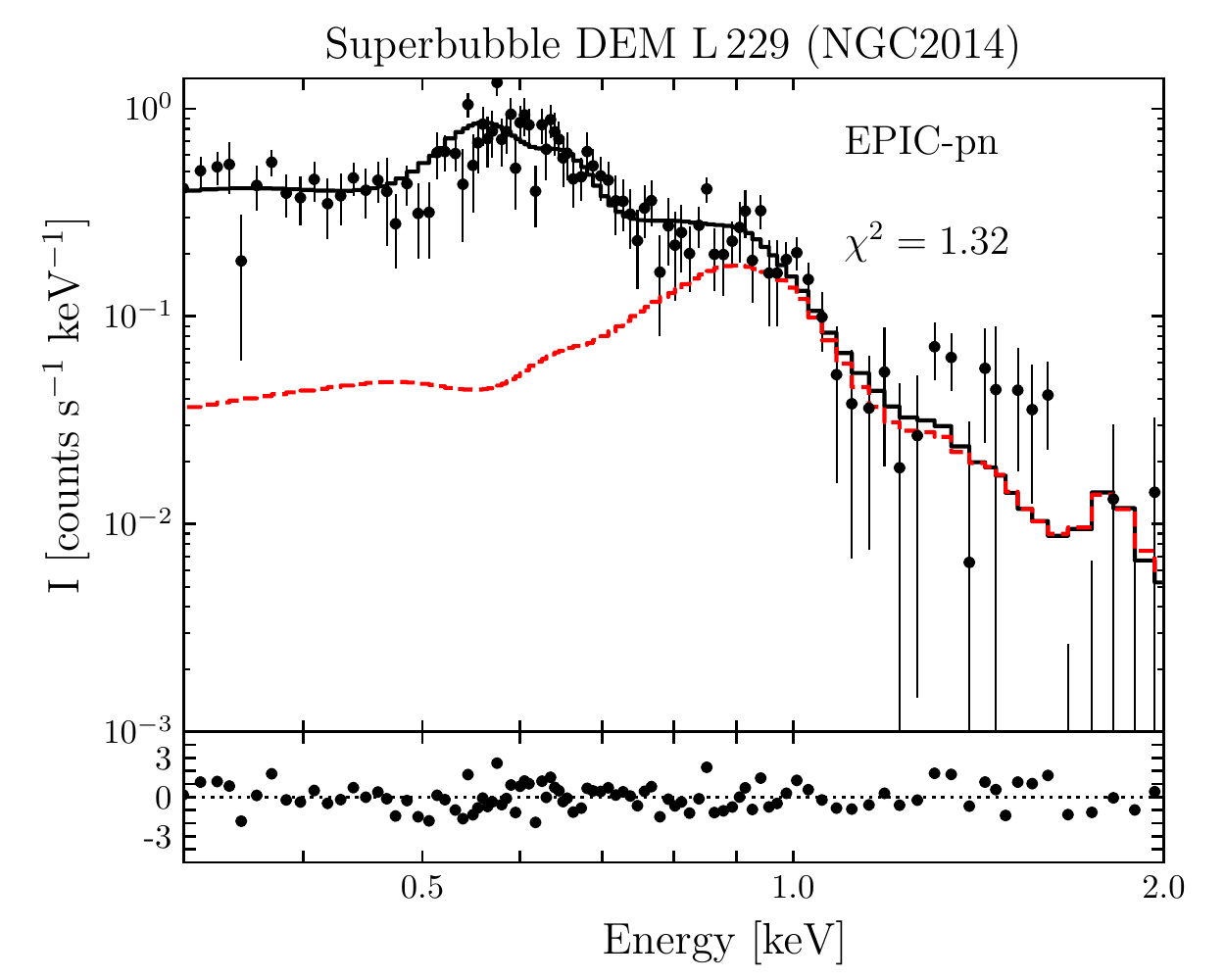}~
\includegraphics[width=0.5\linewidth]{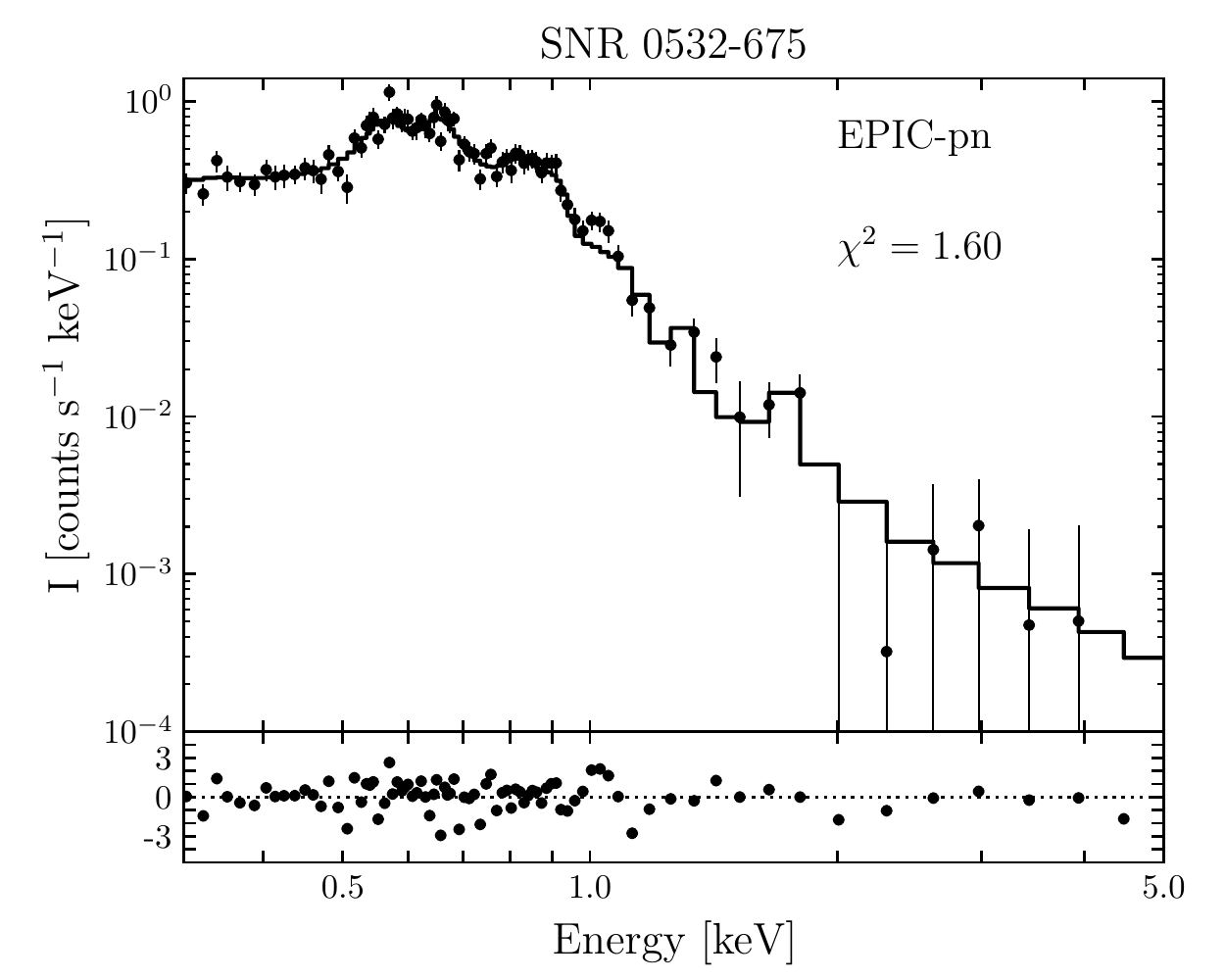}
\caption{Background-subtracted {\it XMM-Newton} EPIC-pn spectra of the
  SB DEM L\,229 and the SNR\,0532-675 in LMC-N\,57. The solid lines
  show the best-fit model. The (red) dashed-line shows the
  contribution of the higher temperature component to the best fit to
  the EPIC-pn data.}
\label{fig:spectra_LOG}
\end{figure*}

\end{document}